\documentclass[prd,twocolumn,showpacs,amsmath,amssymb]{revtex4}
\usepackage{overpic}
\usepackage{graphicx}
\usepackage{epsfig}
\usepackage{dcolumn}
\usepackage{bm}
\usepackage{rotating}
\usepackage{multirow}
\usepackage{subfigure}
\usepackage{hhline}
\usepackage{amssymb}
\usepackage{lineno}
\usepackage[colorlinks,linkcolor=blue,anchorcolor=black,citecolor=blue]{hyperref}

\let\oldequation\equation
\let\oldendequation\endequation

\renewenvironment{equation}
  {\linenomathNonumbers\oldequation}
  {\oldendequation\endlinenomath}

\begin{document}

\title{\boldmath Study of $e^+e^- \to 2(p\bar{p})$ at center-of-mass energies between 4.0 and 4.6 GeV}

\author{
\begin{small}
\begin{center}
M.~Ablikim$^{1}$, M.~N.~Achasov$^{10,c}$, P.~Adlarson$^{67}$, S. ~Ahmed$^{15}$, M.~Albrecht$^{4}$, R.~Aliberti$^{28}$, A.~Amoroso$^{66A,66C}$, Q.~An$^{63,50}$, ~Anita$^{21}$, X.~H.~Bai$^{57}$
% ...
\\
\vspace{0.2cm}
(BESIII Collaboration)\\
\vspace{0.2cm} {\it
$^{1}$ Institute of High Energy Physics, Beijing 100049, People's Republic of China\\
% ...
$^{4}$ Bochum Ruhr-University, D-44780 Bochum, Germany\\
% ...
$^{10}$ G.I. Budker Institute of Nuclear Physics SB RAS (BINP), Novosibirsk 630090, Russia\\
% ...
$^{15}$ Helmholtz Institute Mainz, Johann-Joachim-Becher-Weg 45, D-55099 Mainz, Germany\\
% ...
$^{21}$ Indian Institute of Technology Madras, Chennai 600036, India\\
% ...
$^{28}$ Johannes Gutenberg University of Mainz, Johann-Joachim-Becher-Weg 45, D-55099 Mainz, Germany\\
% ...
$^{50}$ State Key Laboratory of Particle Detection and Electronics, Beijing 100049, Hefei 230026, People's Republic of China\\
% ...
$^{57}$ University of Jinan, Jinan 250022, People's Republic of China\\
% ...
$^{63}$ University of Science and Technology of China, Hefei 230026, People's Republic of China\\
% ...
$^{66}$ University of Turin and INFN, (A)University of Turin , I-10125, Turin, Italy; (B)University of Eastern Piedmont, I-15121, Alessandria, Italy; (C)INFN, I-10125, Turin, Italy\\
$^{67}$ Uppsala University, Box 516, SE-75120 Uppsala, Sweden\\
% ...
$^{c}$ Also at the Novosibirsk State University, Novosibirsk, 630090, Russia\\
% ...
}
\end{center}
\vspace{0.4cm}
\end{small}
 }
\noaffiliation{}

\date{\today}

\begin{abstract}
Using data taken at 23 center-of-mass energies between 4.0 and 4.6~GeV
with the BESIII detector at the BEPCII collider and with a total
integrated luminosity of approximately $15$~fb$^{-1}$, the process
$e^+e^-\to 2(p \bar{p})$ is studied for the first time.  The Born
cross sections for $e^+e^-\to 2(p \bar{p})$ are measured, and no
significant structure is observed in the lineshape.  The baryon pair
($pp$ and $\bar{p}\bar{p}$) invariant mass spectra are consistent with
phase space, therefore no hexaquark or di-baryon state is found.
\end{abstract}

\maketitle
\oddsidemargin -0.2cm
\evensidemargin -0.2cm

\section{Introduction}
Since 2003, a series of charmonium-like states, such as
$X(3872)$~\cite{x3872}, $Y(4260)$~\cite{y42601}, and
$Z_c(3900)$~\cite{zc39001,zc39002}, have been discovered.  The
$Y(4260)$ was first observed by the BABAR experiment via the
initial-state radiative (ISR) process $e^+e^- \to \gamma_{\rm ISR} \pi^+\pi^-
J/\psi$, and it was confirmed by the CLEO~\cite{isry42602} and Belle
experiments~\cite{isry42603}.  In 2017, BESIII reported precise
measurements of the $e^+e^- \to \pi^+\pi^- J/\psi$ cross sections in
the energy region between 3.77 and 4.60~GeV~\cite{ppjpsibes3}. Two
structures are observed with masses of $4222.0\pm3.1\pm1.4$~MeV and
$4320.0\pm10.4\pm7.0$~MeV, and the former, regarded previously as the
$Y(4260)$, is renamed to $Y(4220)$.  The $Y(4220)$ mass is confirmed by
cross-section measurements of $e^+e^-\to \omega \chi_{c0}$~\cite{omegachic0},
$\pi^+\pi^- h_c$~\cite{pphcbes3}, $ \pi^+\pi^-
\psi(3686)$~\cite{pppsip}, and $D^0 D^{*-} \pi^+$~\cite{ddstpi}.

Currently, the known decays of $Y(4220)$ occur only to open or hidden
charm final states.  However, some theories argue that the
charmonium-like states, such as $Y(4220)$, are very likely to also
decay to light hadrons~\cite{hadrocharmonium}.  BESIII has reported
measurements of the cross sections for the light hadron processes of
$e^+e^-\to K_S^0 K^{\pm}\pi^{\mp}\pi^0 (\eta)$~\cite{kskp}, $K_S^0
K^{\pm}\pi^{\mp}$~\cite{kskpi}, $p\bar{n}K_S^0
K^{-}+c.c.$~\cite{kskpn}, and $p\bar{p}\pi^0$~\cite{pppi}, but no hint
of charmless $Y(4220)$ decays has been found. Comprehensive
measurements of the cross sections for $e^+e^-\to light~hadrons$ are
important to search for charmless decays of $Y$ states and to deeply
explore the composition and properties of $Y$ states.

Searches for di-baryon or hexaquark states via $pp$ and $pn$
scattering processes have been carried out in fixed target
experiments.  A resonance $d^*(2380)$ in the isoscalar two-pion fusion
process $pn \to d\pi^0\pi^0$ was observed by WASA~\cite{dp0p0}.  This
state was later confirmed by the other two-pion fusion processes
$pn\to d\pi^+\pi^-$~\cite{dppp0} and $pp\to d\pi^+\pi^0$~\cite{dpppm},
and the two-pion non-fusion processes $pn\to
pp\pi^0\pi^-$~\cite{ppp0pm} and $pn\to pn\pi^0\pi^0$~\cite{pnp0p0}.
However, no experimental information is available in any $e^+e^-$
collision experiment.

In the analysis presented in this paper, we study for the first time
the $e^+ e^- \to 2(p\bar{p})$ process in
the center-of-mass (c.m.) energy ($\sqrt{s}$) region between 4.0 and
4.6~GeV. We search for the $Y(4220)$ structure
by fitting the lineshape of the Born cross sections
measured at these c.m.\ energies.
In addition, we search for a potential structure
similar to the $d^*(2380)$ in the $pp$ and $\bar p \bar p$ mass
spectra.

\section{The BESIII detector and data samples}

The BESIII detector is a magnetic spectrometer~\cite{BESIII} located
at the Beijing Electron Positron Collider~(BEPCII). The cylindrical
core of the BESIII detector consists of a helium-based multilayer
drift chamber (MDC), a plastic scintillator time-of-flight system
(TOF), and a CsI(Tl) electromagnetic calorimeter (EMC), which are all
enclosed in a superconducting solenoidal magnet providing a $1.0$~T
magnetic field. The solenoid is supported by an octagonal flux-return
yoke with resistive plate counter muon identifier modules interleaved
with steel. The acceptance of charged particles and photons is $93$\%
over $4\pi$ solid angle. The charged-particle momentum resolution at
$1~{\rm GeV}/c$ is $0.5\%$, and the specific ionization energy loss
(${\rm d}E/{\rm d}x$) resolution is $6\%$ for the electrons from
Bhabha scattering. The EMC measures photon energies with a resolution
of $2.5\%$ ($5\%$) at $1$~GeV in the barrel (end cap) region. The time
resolution of the TOF barrel part is 68~ps, while that of the end cap
part is 110~ps.

The 23 data sets taken at $\sqrt{s}=4.0\sim4.6$~GeV are used for
this analysis.  The nominal energy of each data set is calibrated by
the di-muon process $e^+e^- \to (\gamma_{\rm
  ISR/FSR})\mu^+\mu^-$~\cite{dimu}, where the subscript ${\rm
  ISR/FSR}$ stands for the initial-state or final-state
radiative process, respectively.  The integrated luminosity $\mathcal{L}$ is
determined using large angle Bhabha events~\cite{rlum}, and the total
integrated luminosity is approximately 15~fb$^{-1}$.

The response of the BESIII detector is modeled with Monte Carlo (MC)
simulations using the software framework BOOST~\cite{boost} based on
GEANT4~\cite{geant4}, which includes the geometry and material
description of the BESIII detectors, the detector response and
digitization models, as well as a database that keeps track of the
running conditions and the detector performance. MC samples are used
to optimize the selection criteria, evaluate the signal efficiency,
and estimate backgrounds.

Inclusive MC samples are generated at different c.m.\ energies to study
the potential backgrounds.  The inclusive MC samples consist of the
production of open-charm processes, the ISR production of vector
charmonium and charmonium-like states, and the continuum processes
incorporated in {\sc kkmc}~\cite{kkmc}. The known decay modes are
modeled with {\sc evtgen}~\cite{besevtgen} using branching fractions
taken from the Particle Data Group (PDG)~\cite{pdg}, and the remaining
unknown decays from the charmonium states with {\sc
  lundcharm}~\cite{lundcharm}.  The FSR from charged final-state
particles are incorporated with {\sc photos}~\cite{photos}.  The
signal MC samples are generated with a phase-space (PHSP) distribution
for the same 23 energy points as data.

\section{data analysis}

For each candidate event, it is required that there are four good
charged tracks.  Two of them must be identified as protons and two as
anti-protons.  The charged particles are required to be within the
acceptance range of $|\cos\theta|<0.93$, where $\theta$ is the polar
angle with respect to the MDC axis. All the charged tracks are
required to originate from the interaction region $R_{xy}<1$~cm and
$|V_{z}|<10$~cm, where $R_{xy}$ and $|V_{z}|$ are the distances of
closest approach of the charged track to the interaction point in the
$xy$-plane and $z$ direction, respectively. For particle
identification (PID), the ${\rm d}E/{\rm d}x$ measured by the MDC and
the TOF information are used to calculate the confidence levels for
the particle hypotheses of pion, kaon and proton.  If the confidence
level for the proton (anti-proton) hypothesis is larger than that for the
other two hypotheses, it is identified as a proton (anti-proton).
The efficiency of PID is $80\%\sim100\%$ as a function of transverse momentum of proton (anti-proton).

A three-constraint ($3$C) kinematic fit imposing three-momentum
conservation under the hypothesis of $e^+e^- \to 2(p\bar{p})$ is
performed for the four candidate charged tracks to suppress background
events.  Since the energy will be used in determining the signal
yield, it is not constrained in the kinematic fit.  The candidate
events with $\chi^2_{\rm 3C}<60$ are kept for further analysis.

\begin{figure}[htbp]
  \begin{center}
    \begin{overpic}[width=0.40\textwidth]
      {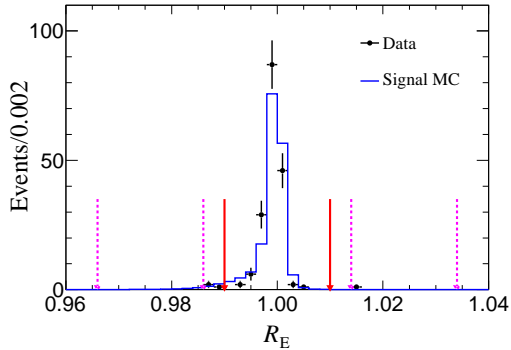}
    \end{overpic}
  \end{center}
  \vspace{-0.8cm}
  \caption{Distributions of $R_E$ of the accepted candidate events
    in data (dots with error bars) and signal MC simulation
    (histogram) from all c.m.\ energies.  The region between the two solid
    red arrows is the signal region, while the region between the two
    dashed pink arrows is the sideband region.  }
  \label{fig:RE}
\end{figure}

The signal yield is determined by a kinematic variable $R_E = E_{\rm
  measure}/E_{\rm cm}$, where $E_{\rm measure}$ is the total energy of
all final particles and $E_{\rm cm}$ is the c.m.\ energy.
Figure~\ref{fig:RE} shows the $R_E$ distribution of the accepted
candidate events.  The signal events concentrate around $1.0$ in the
$R_E$ distribution.  The signal region is defined as the region with
$R_E\in (0.99,1.01)$, while the sideband region is defined as the
region with $R_E\in(0.966,0.986)\cup(1.014,1.034)$.  Studies based on
the inclusive MC samples show that only a few background events
survive at $4.180$~GeV, and they do not form a peak. The
background in the $R_E$ signal region is estimated by the events in
the $R_E$ sideband region multiplied by a scale factor of $0.5$
assuming that the background is flat.  The numbers of events in the
$R_E$ signal region in data and the scaled background yields, which
are obtained by counting, are summarized in the third and fourth
columns of Table~\ref{tab:sec}.

\section{Detection efficiency }

\begin{figure*}[htbp]
  \begin{center}
    \begin{overpic}[width=1\textwidth]
      {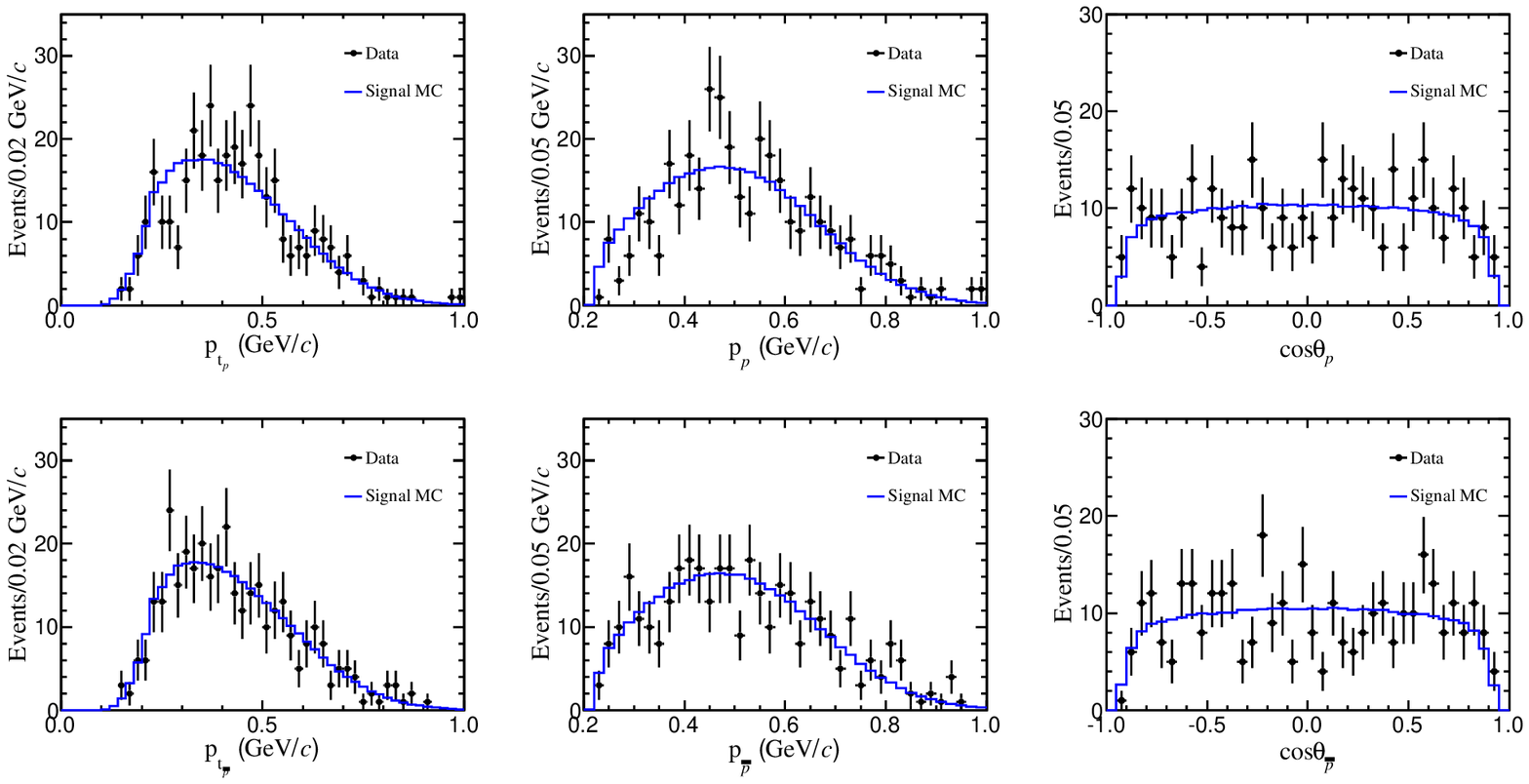}
    \end{overpic}
  \end{center}
  \vspace{-0.8cm}
   \caption{Transverse-momentum distribution (left) and the
     momentum distribution (middle), the polar-angle distribution
     (right) for the proton (top) and anti-proton (bottom) for all data (filled circles with error bars)
     and signal MC simulation (histogram).
     Please note that there are two entries for each event.}
  \label{fig:datamccom}
\end{figure*}

\begin{figure}[htbp]
	\centering
	\includegraphics[width=0.5\textwidth]{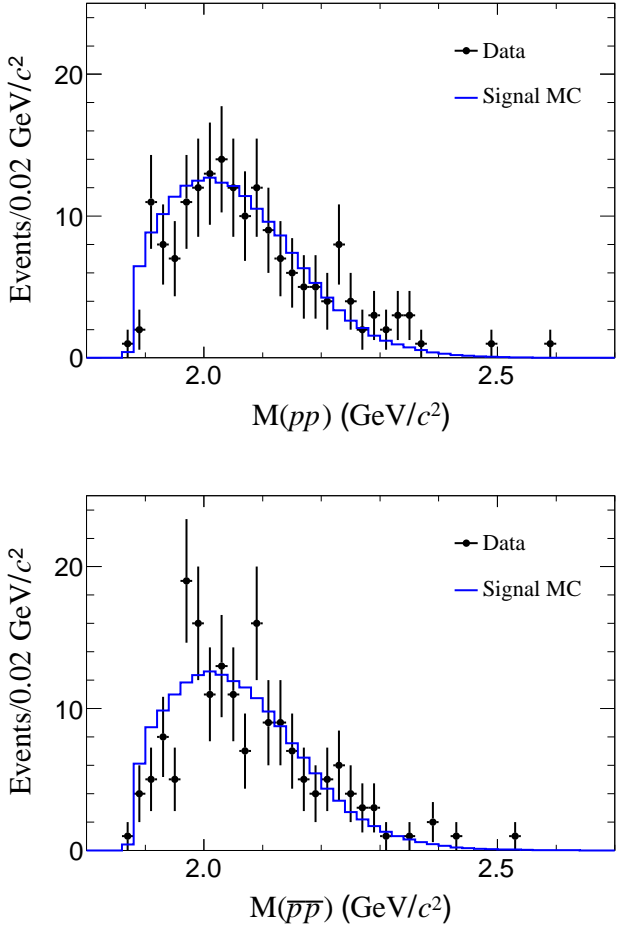}
       \vspace*{-0.8cm}
	\caption{ Invariant-mass distributions of $pp$ (top) and
          $\bar{p}\bar{p}$ (bottom) from all data (filled circles with error bars) and signal MC simulation (histogram).}
	\label{fig:com4}
\end{figure}

Figure~\ref{fig:datamccom} compares momenta, transverse momenta, and polar angle
distributions between accepted candidate events of data and signal MC
samples. Throughout the paper, the data and MC distributions sum
over all c.m.\ energies unless stated
otherwise, while the MC distributions have been weighted by the data
signal yields.  There is good agreement between data and MC
simulation. Therefore, the MC events generated according to PHSP are
used to determine the detection efficiency.  The $pp$ and $\bar p\bar
p$ invariant-mass spectra are shown in Fig.~\ref{fig:com4}, and no
obvious structure is found.

\begin{figure}[htbp]
\centering
\includegraphics[width=0.5\textwidth]{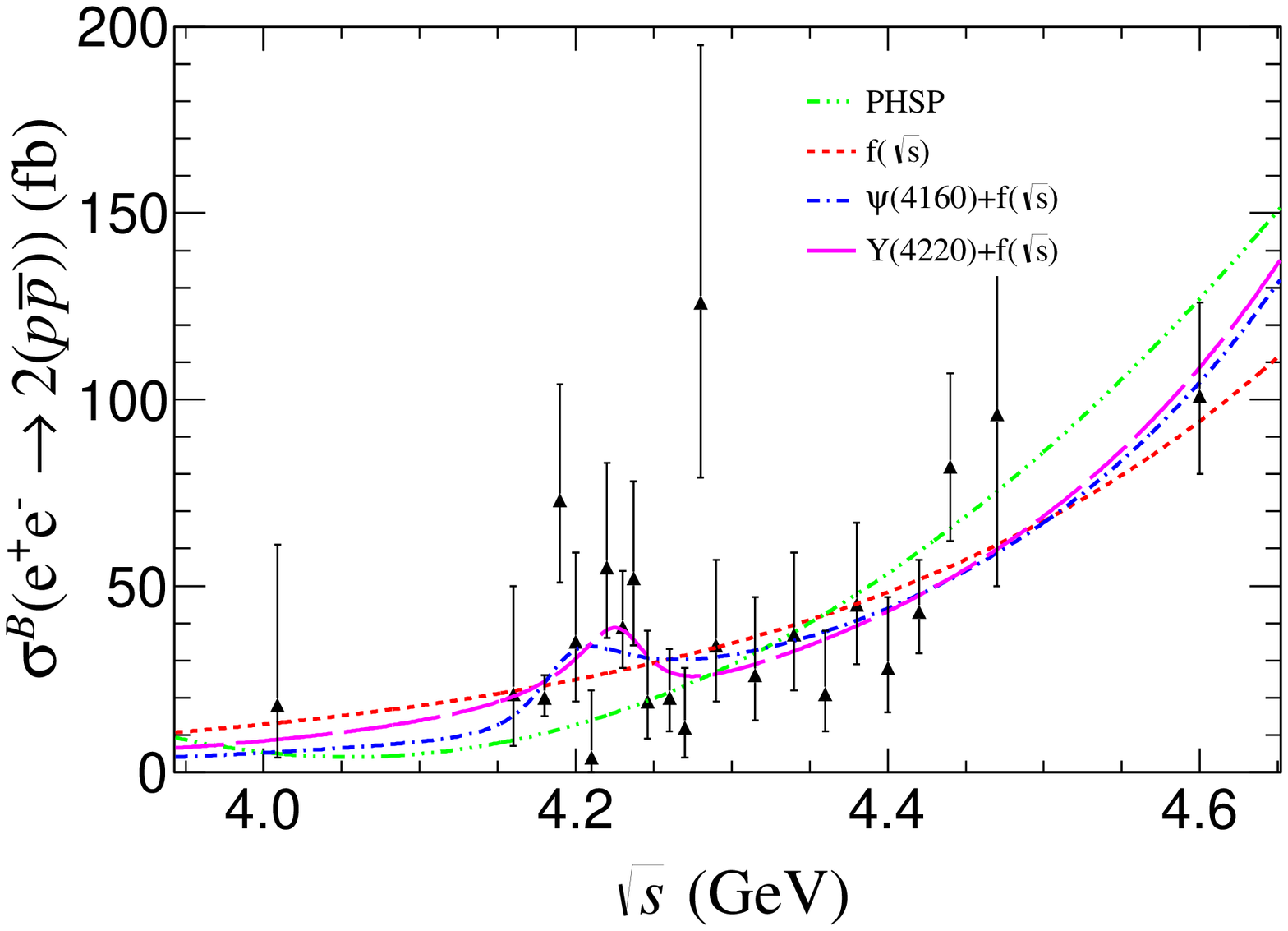}
\vspace*{-0.8cm}
\caption{Born cross sections of the process $e^+e^- \to 2(p\bar{p})$ as a function of c.m.\ energy.
  The data are presented as filled triangles with error bars corresponding to the combination of statistical and systematic uncertainties.
  The lines are fit results to various model assumptions which are described in the text.
}
\label{fig:fitsec}
\end{figure}

The Born cross section of $e^+e^- \to 2(p\bar{p})$ at each c.m.\ energy
is calculated as
\begin{equation} \label{eq:sigmaobs}
\sigma^{\rm Born} = \frac{N^{\rm net}}{\mathcal{L}\times\epsilon \times (1+\delta^{\gamma})\times \frac{1}{\vert 1-\Pi\vert^2}} ,
\end{equation}
where $N^{\rm net}$ is the net number of signal events after
background subtraction, $\mathcal{L}$ is the integrated luminosity of
the data set, $\epsilon$ is the detection efficiency,
$(1+\delta^{\gamma})$ and $\frac{1}{\vert 1-\Pi\vert^2}$ are the
ISR and vacuum polarization correction factors, respectively.

To obtain $(1+\delta^{\gamma})$ and $\frac{1}{\vert
  1-\Pi\vert^2}$, we take the cross section according to the
energy-dependent lineshape of $1/s$ as the initial input, and obtain
the Born cross section by iteration until the efficiencies become
stable at all energies. The difference of
$\epsilon\cdot(1+\delta^{\gamma})$ between the last two iterations is
required to be less than 1\%.  The relevant numbers related to Born
cross section measurement are summarized in Table~\ref{tab:sec}.


\begin{table*}[htbp]
\begin{center}
\caption{The integrated luminosities ($\mathcal{L}$), detection
  efficiencies ($\epsilon$), radiative correction factors
  $(1+\delta^{\gamma})$, vacuum polarization factors
  $(1+\delta^{\nu})$, and the Born cross section ($\sigma^{\rm B}$) at
  different c.m.\ energies ($\sqrt{s}$). The first uncertainties for
  cross sections are statistical and the second are systematic,
  respectively, while those for $N^{\rm obs}$, $N^{\rm bkg}$, $N^{\rm
    net}$ and $\epsilon$ are statistical only.  }
\label{tab:sec}
\begin{tabular}{ccccccccccccc}\toprule
 $\sqrt{s}$~(GeV)   & $N^{\rm obs}_{\rm data}$ & $N^{\rm bkg}_{\rm data}$ & $N^{\rm net}_{\rm data}$  & $\mathcal{L}$~(pb$^{-1}$)   &$\epsilon(\%)$ & $(1+\delta^{\gamma})$   & $\frac{1}{\vert 1-\Pi\vert^2}$  & $\sigma^{\rm Born}$ (fb)   \\\hline
 4.009   &$1.0_{-0.8}^{+2.3}$ &$0.0_{-0.0}^{+0.5}$ &$1.0_{-0.8}^{+2.4}$      &$482.0$   &$13.6 \pm0.1$ &0.8211  &1.0441  &$18_{-14}^{+43}\pm2$  \\
 4.160   &$2.0_{-1.3}^{+2.6}$ &$0.0_{-0.0}^{+0.5}$ &$2.0_{-1.3}^{+2.7}$       &$406.9$  &$26.0 \pm0.1$ &0.8492  &1.0533  &$21_{-14}^{+29}\pm2$  \\
 4.180   &$16.0_{-4.0}^{+5.1}$ &$0.0_{-0.0}^{+0.5}$ &$16.0_{-4.0}^{+5.1}$  &$3194.5$    &$28.6 \pm0.1$ &0.8502	&1.0541   &$20_{-5}^{+6}\pm2$    \\
 4.190   &$10.0_{-3.1}^{+4.3}$ &$0.0_{-0.0}^{+0.5}$ &$10.0_{-3.1}^{+4.3}$      &$523.9$ &$29.3 \pm0.1$ &0.8503  &1.0558  &$73_{-22}^{+31}\pm6$  \\
 4.200   &$5.0_{-2.2}^{+3.4}$ &$0.0_{-0.0}^{+0.5}$ &$5.0_{-2.2}^{+3.4}$     &$525.2$    &$30.0 \pm0.1$ &0.8515  &1.0565  &$35_{-16}^{+24}\pm3$ \\
 4.210   &$1.0_{-0.8}^{+2.3}$ &$0.5_{-0.4}^{+1.2}$ &$0.5_{-0.9}^{+2.6}$    &$517.2$     &$30.7 \pm0.1$ &0.8522  &1.0568  &$3_{-6}^{+18}\pm1$   \\
 4.220   &$8.0_{-2.8}^{+4.0}$ &$0.0_{-0.0}^{+0.5}$ &$8.0_{-2.8}^{+4.0}$      &$513.4$   &$31.4 \pm0.1$ &0.8515  &1.0563  &$55_{-19}^{+28}\pm4$ \\
 4.230   &$12.0_{-3.4}^{+4.6}$ &$0.0_{-0.0}^{+0.5}$ &$12.0_{-3.4}^{+4.6}$    &$1056.4$  &$32.1 \pm0.1$ &0.8529  &1.0564  &$39_{-11}^{+15}\pm3$ \\
 4.237   &$8.0_{-2.8}^{+4.0}$ &$0.0_{-0.0}^{+0.5}$ &$8.0_{-2.8}^{+4.0}$  &$529.1$       &$32.5 \pm0.1$ &0.8527  &1.0555  &$52_{-18}^{+26}\pm4$ \\
 4.246   &$3.0_{-1.6}^{+2.9}$ &$0.0_{-0.0}^{+0.5}$ &$3.0_{-1.6}^{+3.0}$  &$536.3$       &$33.1 \pm0.1$ &0.8535  &1.0555  &$19_{-10}^{+19}\pm2$ \\
 4.260   &$5.0_{-2.2}^{+3.4}$ &$0.0_{-0.0}^{+0.5}$ &$5.0_{-2.2}^{+3.4}$  &$828.4$       &$33.9 \pm0.1$ &0.8543  &1.0536  &$20_{-9}^{+13}\pm2$  \\
 4.270   &$2.0_{-1.3}^{+2.6}$ &$0.0_{-0.0}^{+0.5}$ &$2.0_{-1.3}^{+2.7}$  &$529.7$       &$34.5 \pm0.2$ &0.8545  &1.0530  &$12_{-8}^{+16}\pm1$  \\
 4.280   &$7.0_{-2.6}^{+3.8}$ &$0.0_{-0.0}^{+0.5}$ &$7.0_{-2.6}^{+3.8}$  &$175.2$       &$35.1 \pm0.2$ &0.8545  &1.0530 &$126_{-47}^{+69}\pm9$  \\
 4.290   &$5.0_{-2.2}^{+3.4}$ &$0.0_{-0.0}^{+0.5}$ &$5.0_{-2.2}^{+3.4}$  &$491.5$       &$33.7 \pm0.1$ &0.8541  &1.0527  &$34_{-15}^{+23}\pm3$  \\
 4.315   &$4.0_{-1.9}^{+3.2}$ &$0.0_{-0.0}^{+0.5}$ &$4.0_{-1.9}^{+3.2}$  &$492.1$       &$35.0 \pm0.2$ &0.8554  &1.0522  &$26_{-12}^{+21}\pm2$  \\
 4.340   &$6.0_{-2.4}^{+3.6}$ &$0.0_{-0.0}^{+0.5}$ &$6.0_{-2.4}^{+3.6}$  &$501.1$       &$36.2 \pm0.2$ &0.8557  &1.0508  &$37_{-15}^{+22}\pm3$  \\
 4.360   &$4.0_{-1.9}^{+3.2}$ &$0.0_{-0.0}^{+0.5}$ &$4.0_{-1.9}^{+3.2}$  &$543.9$       &$39.2 \pm0.2$ &0.8556  &1.0511  &$21_{-10}^{+17}\pm2$   \\
 4.380   &$8.0_{-2.8}^{+4.0}$ &$0.0_{-0.0}^{+0.5}$ &$8.0_{-2.8}^{+4.0}$ &$522.8$        &$38.0 \pm0.2$ &0.8560  &1.0513  &$45_{-16}^{+22}\pm4$  \\
 4.400   &$5.0_{-2.2}^{+3.4}$ &$0.0_{-0.0}^{+0.5}$ &$5.0_{-2.2}^{+3.4}$  &$505.0$       &$38.9 \pm0.2$ &0.8558  &1.0510  &$28_{-12}^{+19}\pm3$  \\
 4.420   &$16.0_{-4.0}^{+5.1}$ &$0.0_{-0.0}^{+0.5}$ &$16.0_{-4.0}^{+5.1}$   &$1043.9$   &$39.7 \pm0.2$ &0.8552  &1.0524  &$43_{-11}^{+14}\pm4$   \\
 4.440   &$17.0_{-4.1}^{+5.2}$ &$0.0_{-0.0}^{+0.5}$ &$17.0_{-4.1}^{+5.2}$ &$568.1$      &$40.4 \pm0.2$ &0.8548  &1.0537  &$82_{-20}^{+25}\pm6$  \\
 4.470   &$4.0_{-1.9}^{+3.2}$ &$0.0_{-0.0}^{+0.5}$ &$4.0_{-1.9}^{+3.2}$  &$111.1$       &$41.5 \pm0.2$ &0.8547  &1.0548  &$96_{-46}^{+77}\pm7$  \\
 4.600   &$24.0_{-4.9}^{+6.0}$ &$0.0_{-0.0}^{+0.5}$ &$24.0_{-4.9}^{+6.0}$ &$586.9$      &$45.0 \pm0.2$ &0.8551  &1.0546  &$101_{-21}^{+25}\pm8$   \\\hline

 \end{tabular}

\end{center}
\end{table*}

\section{systematic uncertainty}

The uncertainty in the measurement of the integrated luminosity of the
data set is $1.0$\%~\cite{rlum}.  The uncertainties of the tracking
and PID efficiencies have been studied with high purity control
samples of $J/\psi\to p\bar{p}\pi^+\pi^-$ and $\psi(3686)\to
\pi^+\pi^-J/\psi\to
\pi^+\pi^-p\bar{p}$~\cite{prd91112004,prd99031101}.  The differences
of the tracking and PID efficiencies between data and MC simulation in
different transverse momentum and momentum ranges are obtained
separately. The averaged differences for tracking (PID) efficiencies
that are re-weighted by the proton and anti-proton momenta of the
signal MC events, $0.5$\% (1.3\%) per proton and $1.0$\% (1.8\%) per
anti-proton, are assigned as the tracking (PID) systematic
uncertainties. Adding them linearly gives our estimate of the total systematic
uncertainty of the tracking (PID) efficiency for all charged tracks resulting in
$3.0$\% (6.2\%).

To determine the systematic error related to uncertainties in the signal window, we
define a ratio which is the number of net signal events in the signal
window obtained by counting to the number of signal events obtained by
fitting to the $R_E$ spectra. For data, the combined $R_E$ spectra are
fitted with a MC-derived shape convolved with a Gaussian to describe
the signal shape, while the background shape is described by a
first-order Chebychev polynomial.  For MC-simulated data at each energy
point, the ratio is similarly obtained.  The relative differences
of the ratio between data and MC simulations range from $0.5$\% to
$1.4$\% depending on the energy point and are taken as the uncertainties.

To obtain reliable detection efficiencies, the Born cross sections
input in the generator have been iterated until the
$(1+\delta^{\mathit{r}})\cdot\epsilon$ values converge.  The
differences of $(1+\delta^{\mathit{r}})\cdot\epsilon$ between the last
two iterations in the cross section measurements, which range from
0.0\% to 1.0\%, are taken as the systematic uncertainties due to the
ISR correction factor.

The systematic uncertainty from the kinematic fit is estimated by
changing the $\chi^2_{\rm 3C}$ requirement by $\pm 15$. The largest
changes of the cross sections compared to the nominal $\chi^2_{\rm
  3C}$ requirement range from $0.3$\% to $0.7$\% and are taken as
the corresponding uncertainties.
The total systematic uncertainty is determined, to be $7.0$\% to
$7.1$\%, by summing the individual values in quadrature under the
assumption that all the sources are independent.

\section{\boldmath Fit to the Born cross sections}

We fit to the Born cross sections under four assumptions with the
least-square method~\cite{leastchi2}. In order to describe purely continuum
production, the first cases is
based on a simple four-body energy-dependent PHSP
lineshape~\cite{phsp}. And the second case is based on an exponential
function~\cite{barbarexpf,besexpf}. The exponential function is constructed as
\begin{equation} \sigma^{\rm Born}(s) = \frac{1}{s} \times e^{-p_{\rm
    0}(\sqrt{s}-M_\text{th})} \times p_{\rm 1},
\end{equation}
where $p_{0}$ and
$p_{1}$ are free parameters, and $M_\text{th}$ is $(2m_{p}+2m_{\bar{p}})$.
The third (fourth) case is based on an exponential function for
continuum production plus the well-established charmonium state $\psi(4160)$
(charmonium-like state $Y(4220)$) for resonance production. For the
latter two cases, the light-hadron production is
described as
\begin{equation} \sigma^{\rm Born}(s) = \left|{\rm BW}( \sqrt{{\it
    s}})e^{i\phi} + \sqrt{f(\sqrt{{\it s}})}\right|^2 ,
\end{equation}
where
$\sqrt{f(\sqrt{s})}$ denotes the chosen continuum production
amplitude, the resonance amplitude is described by a relativistic
Breit-Wigner amplitude ${\rm BW}(\rm \sqrt{{\it
    s}})=\frac{\sqrt{12\pi\Gamma^{ee}\Gamma^{tot}}}{{\it s}-M^{2}+{\it
    i}M\Gamma^{tot}}$, and $\phi$ is the phase angle between the two
amplitudes.  Moreover, $\rm M$, $\Gamma^{ee}$ and $\Gamma^{\rm tot}$
are the mass, partial width to $e^+e^-$ and total width of the assumed
resonance, and the values are taken from the PDG~\cite{pdg}, which are
$\rm M =4.191$ $(4.23)$~GeV/$c^{2}$ and $\Gamma^{\rm tot}=70$
$(55)$~MeV for  $\psi(4160)$ ($Y(4220)$).
The fit results are shown in Fig.~\ref{fig:fitsec}.

The goodness-of-fit is $\chi^2/{\rm NDF}=1.9$, 1.2, 1.3, and 1.2 for
the four fit scenarios, respectively. Here, ${\rm NDF}$ is
the number of degrees of freedom.  The large goodness-of-fit for the
first case implies that it is less preferable to describe the $e^+e^- \rightarrow 2(p\bar{p})$ process by a simple four-body energy-dependent PHSP assumption,
while it can be by the exponential function. The statistical
significances of the resonances are estimated by comparing the
change of $\chi^{2}$ with and without the resonance
and taking the change of degrees of freedom into account. The
statistical significance is $0.83\sigma$ ($1.69\sigma$) for
$\psi(4160)$ ($Y(4220)$), which indicates it is unnecessary to
include the $\psi(4160)$ or $Y(4220)$ resonance. This could also imply
that the $\psi(4160)$ and $Y(4220)$ disfavor a decay to $2(p\bar{p})$.

\section{summary}

In conclusion, the process of $e^+e^- \to 2(p\bar{p})$ is studied at
23 c.m.\ energies in the region from 4.0 to 4.6~GeV. The Born cross
sections at the different c.m.\ energies are measured, and the
lineshape can be generally described by an empirical exponential
function. The significances for possible contributions by a
$\psi(4160)$ or $Y(4220)$ resonance are small, namely $0.83\sigma$ and $1.69\sigma$,
respectively. With the present statistics it is impossible to draw
any conclusion whether there are actual resonances or structures in this
lineshape.

The baryon-pair invariant-mass spectra are in good agreement
with phase space, and no hexaquark or di-baryon state is found with
the currently available statistics.

\section{ACKNOWLEDGMENT}
The BESIII collaboration thanks the staff of BEPCII and the IHEP computing center for
their strong support. This work is supported in part by National Key Basic Research
Program of China under Contract No. 2015CB856700; National Natural Science Foundation
of China (NSFC) under Contracts Nos. 11975118, 11625523, 11635010, 11735014, 11822506, 11835012, 11935015, 11935016, 11935018, 11961141012; the Natural Science Foundation of Hunan Province under Contract No. 2019JJ30019; the Chinese Academy of Sciences (CAS) Large-Scale Scientific Facility Program; Joint Large-Scale Scientific Facility Funds of the NSFC and CAS under Contracts Nos. U1732263, U1832207; CAS Key Research Program of Frontier Sciences under Contracts Nos. QYZDJ-SSW-SLH003, QYZDJ-SSW-SLH040; 100 Talents Program of CAS; INPAC and Shanghai Key Laboratory for Particle Physics and Cosmology; ERC under Contract No. 758462; German Research Foundation DFG under Contracts Nos. Collaborative Research Center CRC 1044, FOR 2359; Istituto Nazionale di Fisica Nucleare, Italy; Ministry of Development of Turkey under Contract No. DPT2006K-120470; National Science and Technology fund; STFC (United Kingdom); The Knut and Alice Wallenberg Foundation (Sweden) under Contract No. 2016.0157; The Royal Society, UK under Contracts Nos. DH140054, DH160214; The Swedish Research Council; U. S. Department of Energy under Contracts Nos. DE-FG02-05ER41374, DE-SC-0012069.

\end{document}